# Repetitive Patterns in Rapid Optical Variations in the Nearby Black-hole Binary V404 Cygni

Mariko Kimura[1], Keisuke Isogai[1], Taichi Kato[1], Yoshihiro Ueda[1], Satoshi Nakahira[2], Megumi Shidatsu[3], Teruaki Enoto[1,4], Takafumi Hori[1], Daisaku Nogami[1], Colin Littlefield[5], Ryoko Ishioka[6], Ying-Tung Chen[6], Sun-Kun King[6], Chih-Yi Wen[6], Shiang-Yu Wang[6], Matthew J. Lehner[6,7,8], Megan E. Schwamb[6], Jen-Hung Wang[6], Zhi-Wei Zhang[6], Charles Alcock[8], Tim Axelrod[9], Federica B. Bianco[10], Yong-Ik Byun[11], Wen-Ping Chen[12], Kem H. Cook[6], Dae-Won Kim[13], Typhoon Lee[6], Stuart L. Marshall[14], Elena P. Pavlenko[15], Oksana I. Antonyuk[15], Kirill A. Antonyuk[15], Nikolai V. Pit[15], Aleksei A. Sosnovskij[15], Julia V. Babina[15], Aleksei V. Baklanov[15], Alexei S. Pozanenko[16,17], Elena D. Mazaeva[16], Sergei E. Schmalz[18], Inna V. Reva[19], Sergei P. Belan[15], Raguli Ya. Inasaridze[20], Namkhai Tungalag[21], Alina A. Volnova[16], Igor E. Molotov[22], Enrique de Miguel[23,24], Kiyoshi Kasai[25], William Stein[26], Pavol A. Dubovsky[27], Seiichiro Kiyota[28], Ian Miller[29], Michael Richmond[30], William Goff[31], Maksim V. Andreev[32,33], Hiromitsu Takahashi[34], Naoto Kojiguchi[35], Yuki Sugiura[35], Nao Takeda[35], Eiji Yamada[35], Katsura Matsumoto[35], Nick James[36], Roger D. Pickard[37,38], Tamás Tordai[39], Yutaka Maeda[40], Javier Ruiz[41,42,43], Atsushi Miyashita[44], Lewis M. Cook[45], Akira Imada[46] & Makoto Uemura[47]

[1]Department of Astronomy, Graduate School of Science, Kyoto University, Oiwakecho, Kitashirakawa, Sakyo-ku, Kyoto 606-8502, Japan

[2]JEM Mission Operations and Integration Center, Human Spaceflight Technology Directorate, Japan Aerospace Exploration Agency, 2-1-1 Sengen, Tsukuba, Ibaraki 305-8505, Japan

[3]MAXI team, RIKEN, 2-1 Hirosawa, Wako, Saitama 351-0198, Japan

[4]The Hakubi Center for Advanced Research, Kyoto University, Kyoto 606-8302, Japan

[5]Astronomy Department, Wesleyan University, Middletown, CT 06459 USA

[6]Institute of Astronomy and Astrophysics, Academia Sinica, 11F of Astronomy-Mathematics Building, AS/NTU. No.1, Sec. 4, Roosevelt Rd, Taipei 10617, Taiwan (R.O.C.)

[7]Department of Physics and Astronomy, University of Pennsylvania, 209 S. 33rd St., Philadelphia, PA 19125, USA

[8]Harvard-Smithsonian Center for Astrophysics, 60 Garden St, Cambridge, MA 02138, USA

[9]Steward Observatory, University of Arizona, Tucson, Arizona 85721, USA

[10]Center for Cosmology and Particle Physics, New York University, 4 Washington Place, New York, New York 10003, USA

[11]Department of Astronomy and University Observatory, Yonsei University, Seoul 120-749, Korea

[12]Institute of Astronomy and Department of Physics, National Central University, Chung-Li 32054, Taiwan (R.O.C.)

[13]Max Planck Institute for Astronomy, Königstuhl 17, 69117 Heidelberg, Germany

[14]Kavli Institute for Particle Astrophysics and Cosmology (KIPAC), Stanford University, 452 Lomita Mall, Stanford, California 94309, USA

[15]Crimean Astrophysical Observatory, 298409, Nauchny, Republic of Crimea

[16]Space Research Institute, Russian Academy of Sciences, 117997 Moscow, Russia

[17]National Research Nuclear University MEPhI (Moscow Engineering Physics Institute), Moscow, Russia

[18]Leibniz Institute for Astrophysics, Potsdam, Germany

[19]Fesenkov Astrophysical Institute, Almaty, Republic of Kazakhstan

[20]Kharadze Abastumani Astrophysical Observatory, Ilia State University, Tbilisi, Georgia

[21]Center of Astronomy and Geophysics Mongolian Academy of Sciences, Ulaanbaatar, Mongolia

[22]Keldysh Institute of Applied Mathematics, Russian Academy of Sciences, Moscow, Russia

[23]Departamento de Física Aplicada, Facultad de Ciencias Experimentales, Universidad de Huelva, 21071 Huelva, Spain

[24]Center for Backyard Astrophysics, Observatorio del CIECEM, Parque Dunar, Matalascañas, 21760 Almonte, Huelva, Spain

[25]Baselstrasse 133D, CH-4132 Muttenz, Switzerland

[26]6025 Calle Paraiso, Las Cruces, New Mexico 88012, USA

[27]Vihorlat Observatory, Mierova 4, Humenne, Slovakia

[28]Variable Star Observers League in Japan (VSOLJ), 7-1 Kitahatsutomi, Kamagaya, Chiba 273-0126

[29]Furzehill House, Ilston, Swansea, SA2 7LE, UK

[30]Physics Department, Rochester Institute of Technology, Rochester, New York 14623, USA

[31]American Association of Variable Star Observers (AAVSO), 13508 Monitor Lane Sutter Creek, CA 95685

[32]Institute of Astronomy, Russian Academy of Sciences, 361605 Peak Terskol, Kabardino-Balkaria, Russia

[33]International Center for Astronomical, Medical and Ecological Research of National Academy of Sciences of Ukraine (NASU), 27 Akademika Zabolotnoho str., 03680 Kiev, Ukraine

[34]Department of Physical Science, School of Science, Hiroshima University, 1-3-1 Kagamiyama, Higashi-Hiroshima, Hiroshima 739-8526, Japan

[35]Osaka Kyoiku University, 4-698-1 Asahigaoka, Kashiwara, Osaka 582-8582, Japan

[36]1 Tavistock Road, Chelmsford, Essex, CM1 6JL, UK

[37]The British Astronomical Association, Variable Star Section (BAA VSS), Burlington House, Piccadilly, London, W1J 0DU, UK

[38]3 The Birches, Shobdon, Leominster, Herefordshire, HR6 9NG, UK

[39]Polaris Observatory, Hungarian Astronomical Association, Laborc utca 2/c, 1037 Budapest, Hungary

[40]112-14 Kaminishiyama-machi, Nagasaki, Nagasaki 850-0006, Japan

[41]Observatorio de Cantabria, Ctra. de Rocamundo s/n, Valderredible, Cantabria, Spain

[42]Instituto de Fisica de Cantabria (CSIC-UC), Avenida Los Castros s/n, E-39005 Santander, Cantabria, Spain

[43]Agrupacion Astronomica Cantabra, Apartado 573, 39080, Santander, Spain

[44] Seikei Meteorological Observatory, Seikei High School, Kichijoji-kitamachi 3-10-13, Musashino, Tokyo 180-8633, Japan

[45]Center for Backyard Astrophysics (Concord), 1730 Helix Ct. Concord, California 94518, USA

[46]Kwasan and Hida Observatories, Kyoto University, Kitakazan-Ohmine-cho, Yamashina-ku, Kyoto, Kyoto 607-8471, Japan

[47]Hiroshima Astrophysical Science Center, Hiroshima University, Kagamiyama 1-3-1, Higashi-Hiroshima, Hiroshima 739-8526, Japan

**How black holes accrete surrounding matter is a fundamental, yet unsolved question in astrophysics. It is generally believed that when the mass accretion rate approaches the critical rate (Eddington limit), thermal instability occurs in the inner disc, causing repetitive patterns[1] of violent X-ray variability (oscillations) on timescales of minutes to hours. In fact, such oscillations have been observed only in high mass accretion rate sources, like GRS 1915+105[2,3]. These phenomena are thought to have distinct physical origins from X-ray or optical variations with much smaller amplitudes and faster ($\lesssim$10 sec) timescales often observed in other black hole binaries (e.g., XTE J1118+480[4] and GX 339−4[5]). Here we report an extensive multi-colour optical photometric dataset of V404 Cygni, an X-ray transient[6] containing a black hole of nine solar masses[7] at a distance of 2.4 kpc[8]. Our data show that optical oscillations on timescales of 100 sec to 2.5 hours can occur at mass accretion rates >10 times lower than previously thought[1]. This suggests that the accretion rate is not the critical parameter for inducing inner disc instabilities. Instead, we propose that a long orbital period is a key condition for these phenomena, because the outer part of the large disc in binaries with long orbital periods will have surface densities too low to maintain the sustained mass accretion to the inner part of the disc. The lack of sustained accretion – not the actual rate – would then be the critical factor causing violent oscillations in long-period systems.**

V404 Cyg, which was originally discovered as a nova in 1938 and detected by the GINGA satellite in 1989[9], underwent an outburst in 2015 June after 26 years of dormancy. At 18:31:38 on June 15 (15.77197 Universal Time (UT)), Swift/Burst Alert Telescope (BAT) initially detected this outburst as a possible gamma-ray burst[10]. The outburst was also detected by the Monitor of All-sky X-ray Image (MAXI) instrument on June 16.783 UT[11].

Following these detections, we started a world-wide photometric campaign (Extended Data Tables 1 and 2, Sec. 1 of Methods) partly within the Variable Star Network (VSNET) Collaboration, and collected extensive sets of multi-colour optical photometric data consisting of >85,000 points. Our dataset also includes early observations with the Taiwanese-American Occultation Survey (TAOS) starting on June 15, 18:34:07 UT, 2 min 29 sec after the Swift/BAT trigger[12] (Extended Data Tables 1 and 2, Sec. 1 of Methods on the VSNET collaboration team and TAOS). Some weak activity started approximately 1,000 s before the Swift/BAT trigger[13]. The same activity above 80 keV was also detected by the active anti-coincidence shield (ACS) of Spectrometer on INTEGRAL (SPI) telescope of INTEGRAL observatory in the same time intervals (P. Minaev, private communication).

Our observations immediately indicated that large-amplitude short-term variations on timescales of $\sim$100 s to $\sim$2.5 hours were already present, starting less than three minutes after the Swift/BAT trigger. In Fig. 1 and Extended Data Fig. 1, we show the overall optical multi-colour light curves. The overall trend of the light curves can be divided into three stages: (1) gradual rise during BJD (Barycentric Julian Day) 2,457,189–2,457,194.5 (brightening by 1 mag $d^{-1}$ on average), (2) the plateau during BJD 2,457,194.5–2,457,200.0, and (3) rapid fading during BJD 2,457,200.0–2,457,203.3 (fading on average by 2.5 mag $d^{-1}$). Short-term variations with amplitudes varying between 0.1–2.5 mag were observed throughout the outburst and consisted of characteristic structures such as recurrent sudden dips from a peak (Fig. 1).

Moreover, fluctuations similar in shapes to the unique X-ray variations of the enigmatic black hole binary GRS 1915+105[2] are present in the optical light curve of V404 Cyg (Fig. 2). The patterns in the X-ray light curve of GRS 1915+105 have been classified into at least 12 classes on the basis of their flux and colour characteristics[3]. Repeating structures like these had not been observed in optical wavelengths prior to the 2015 outburst of V404 Cyg. The variations that we observed can be divided into two characteristic classes: (1) dip-type oscillations (repetitions of a gradual rise followed by a sudden dip, sometimes with accompanying spikes on timescales of $\sim$45 min–$\sim$2.5 hours; Fig. 2a, b, and c) and (2) heartbeat-type oscillations (rhythmic small spikes with short periods of $\sim$5 min; Fig. 2d). Although rapid optical variations have been detected in the black hole binary V4641 Sgr, the variations are stochastic with no indication of regular patterns[14]. The variations we found in V404 Cyg at optical wavelengths were regular and similar in shape to those in GRS 1915+105, although the interval between dips is about 5 times larger in V404 Cyg than in GRS 1915+105.

Using X-ray data from Swift/X-ray Telescope (XRT), we compared simultaneous optical and X-ray light curves (Fig. 3). When both X-ray and optical data showed strong short-term variations, the correlations were generally good, although the X-ray flux variations are much larger than the optical ones. The good correlation indicates that both X-ray and optical observations recorded the same phenomena (see also Sec. 2 in Methods and Extended Data Fig. 2). Spectral analyses of the simultaneous X-ray data (Sec. 3 in Methods and Extended Data Fig. 3) indicate that there was no tendency for increased absorption when the X-ray flux decreased, suggesting that these dips do not originate in absorption. In some epochs, we found evidence for heavy obscuration as found in the GINGA data during the 1989 outburst[15]; however this is not related to dip-type variations. We can thus infer that the short-term fluctuations directly reflect variations in radiation from the accretion disc or its associated structures. Detailed analyses of the typical simultaneous broad-band spectral energy distribution (SED) (Sec. 8 of Methods and Extended Data Fig. 6) show that the majority of the optical flux is most likely produced by reprocessing of X-ray irradiation in the disc.

In GRS 1915+105, it has been proposed that the observed variability is caused by limit-cycle oscillations in the inner accretion disc due to the Lightman-Eardley viscous instability[16], which can explain a slow rise in brightness (mass accumulation) followed by a sudden drop (accretion to the black hole). Such a model assumes that the black hole is accreting mass nearly at the Eddington rate, which is supported by observations of GRS 1915+105[17]. Similar types of X-ray variability have also been detected in the black hole binary IGR J17091−3624[18], whose Eddington rate is unknown because both the mass and the distance are uncertain.

In V404 Cyg, however, the accurate determination of the distance based on a parallax measurement[8] and the dynamical mass determination[7] enable us to conclude from our 2015 data that the black hole in this system was accreting at much lower rate than the Eddington rate most of the time. During the period when GRS 1915+105-type variations in the optical light curves were recorded in V404 Cyg, its bolometric luminosity, averaged over an interval longer than the period of oscillation, spanned a wide range, from $\sim$0.01 $L_{Edd}$ to $\sim$0.4 $L_{Edd}$ (where $L_{Edd}$ is the Eddington luminosity for a nine solar-mass black hole), as estimated from the hard X-ray flux and SED (Fig. 4 and Sec. 9 in Methods). Remarkably, the dip-type oscillations were observed at mean bolometric luminosity of $\sim$0.015

$L_{Edd}$, ~0.07 $L_{Edd}$, and ~0.06 $L_{Edd}$ during BJD 2,457,191.35–2,457,191.60, BJD 2,457,192.34–2,457,192.70, and BJD 2,457,200.60–2,457,200.76, respectively.

It is also worth noting that a typical dip similar to those seen in GRS 1915+105 was observed just 3 min after the first detection of this outburst (Extended Data Fig. 1b). These facts suggest that the accretion rate is not the critical parameter for inducing these oscillations. Our results imply that there is a novel type of disc instability different from the known dwarf-nova type[19] or Lightman-Eardley type[16] instability.

We point out that black hole binaries showing large-amplitude, short-term variations either in X-ray or optical bands have long orbital periods. (33.9 d in GRS 1915+105[20], ~4 d in IGR J17091−3624[21], 6.5 d in V404 Cyg[22], and 2.8 d in V4641 Sgr[23]; Sec. 4 in Methods and Extended Data Table 3 for a comparison of these objects), reinforcing the earlier suggestion[24]. It has been proposed that the accretion disc in a system with a long orbital period suffers from instabilities in the vertical structure of the disc, and hence the disc beyond this radius of instability may never build up[15,25]. Our SED modeling of this outburst, however, requires a disc having a large radius ($\gtrsim 1.7 \times 10^{12}$ [cm]), even considering the uncertainty of the interstellar reddening, particularly to account for the ultraviolet flux. This result implies that the disc extended up to distances close to the maximum achievable radius (Sec. 8 in Methods). This radius is consistent with the short-term optical variations significantly detected below 0.01 Hz (Sec. 6 in Methods) and the time lag of ~1 min between the X-ray and optical light curves (Fig. 3 and Extended Data Fig. 2) if we assume that the optical light mainly comes from reprocessed X-rays (Sec. 6 and Extended Data Fig. 5 in Methods). We note that synchrotron emission has been proposed to be the origin of the short-term and large-amplitude fluctuations in case of V4641 Sgr[14]. The optical polarization of V404 Cyg, however, did not show evidence of significant variations during the 2015 outburst[26,27]. This fact disfavours synchrotron emission as the origin of the short-term variations.

Outbursts of X-ray transients are thought to be triggered by the dwarf-nova type instability: once the surface density at some radius reaches the critical density ($\Sigma_{crit}$) after continuous mass transfer from the secondary star, thermal instability occurs and the disc undergoes an outburst[19]. In systems with long orbital periods, it is difficult for surface densities in the outer disc to reach $\Sigma_{crit}$, which is roughly proportional to the radius[28]. As a result, thermal instability in the inner part of the disc occurs more easily and governs the outburst behaviour[29]. This is probably the reason why long-period systems behave differently than short-period "classical" X-ray transients. In fact, our estimate of the disc mass ($5 \times 10^{25}$ [g]) accreted during the 2015 outburst is far smaller than the mass ($2 \times 10^{28}$ [g]) of a fully built-up disc in quiescence (Sec. 5 in Methods and Extended Data Fig. 4). These values indicate that the surface density was well below the $\Sigma_{crit}$ required to induce thermal instability in most parts of the disc at the onset of the present outburst. Once the X-ray outburst started in the inner region, hydrogen atoms in the outer part of the disc were ionized and "passively" maintained in the hot state as long as the X-ray illumination continued. This explains the large optical fluxes as observed[28]. The rapid decay observed in the 2015 outburst of V404 Cyg may reflect the lack of the exponential decay in long-period systems as theoretically predicted[30]. Because the surface densities in the rest of the disc were too low to sustain the outburst by viscous diffusion[19], only the inner part of the disc was responsible for the dynamics of the present outburst, as inferred from the rapid fading from the outburst (Sec. 7 in Methods). We infer that, in outbursts of IGR J17091−3624[18, 21] and the 1938 outburst of V404 Cyg (Sec. 5 and Extended Data Fig. 4 in Methods), the radius of the active disc is larger, which explains why the duration of those events is longer than that of the 2015 outburst of V404 Cyg.

**Acknowledgements** We acknowledge the variable star observations from the AAVSO International Database contributed by observers worldwide and used in this research. We also thank the INTEGRAL groups for making the products of the ToO data public online at the INTEGRAL Science Data Centre. Work at ASIAA was supported in part by the thematic research program AS-88-TP-A02. A.S.P., E.D.M. and A.A.V. are grateful to Russian Science Foundation (grant 15-12-30016) for support. R.Ya.I. is grateful to the grant RUSTAVELI FR/379/6-300/14 for a partial support. We thank Dr. Hiroyuki Maehara, Mr. Hidehiko Akazawa, Mr. Kenji Hirosawa, and Mr. Josep Lluis for their optical observations. This work was supported by a Grant-in-Aid "Initiative for High-Dimensional Data-Driven Science through Deepening of Sparse Modeling" from the Ministry of Education, Culture, Sports, Science and Technology (MEXT) of Japan (25120007 TK) and (26400228 YU).

**Author Contributions** M.K. led the campaign, performed optical data analysis and compiled all optical data. K.I. and A.I. performed optical data analysis. T.K., Y.U., D.N. and M.U. contributed to science discussions. S.N., M.S., T.E., T.H. and H.T. performed X-ray data analysis. Other authors than those mentioned above performed optical observations. M.K., K.I., T.K., Y.U., S.N., T.E., M.S., and A.I. wrote the manuscript. T.K., Y.U., and D.N. supervised this project. M.K., K.I., T.K., Y.U., T.E., M.S., D.N., C.L., R.I., M.J.L., D.B.B., D.K., E.P.P., A.S.P., I.E.M., M.R., E.M., W.S., S.K., L.M.C., A.I., and M.U. improved the manuscript. All authors have read and approved the manuscript.

**Competing Interests** The authors declare that they have no competing financial interests.

**Author Information** Correspondence and requests for materials should be addressed to M.K. (mkimura@kusastro.kyoto-u.ac.jp).

**Main figures:**

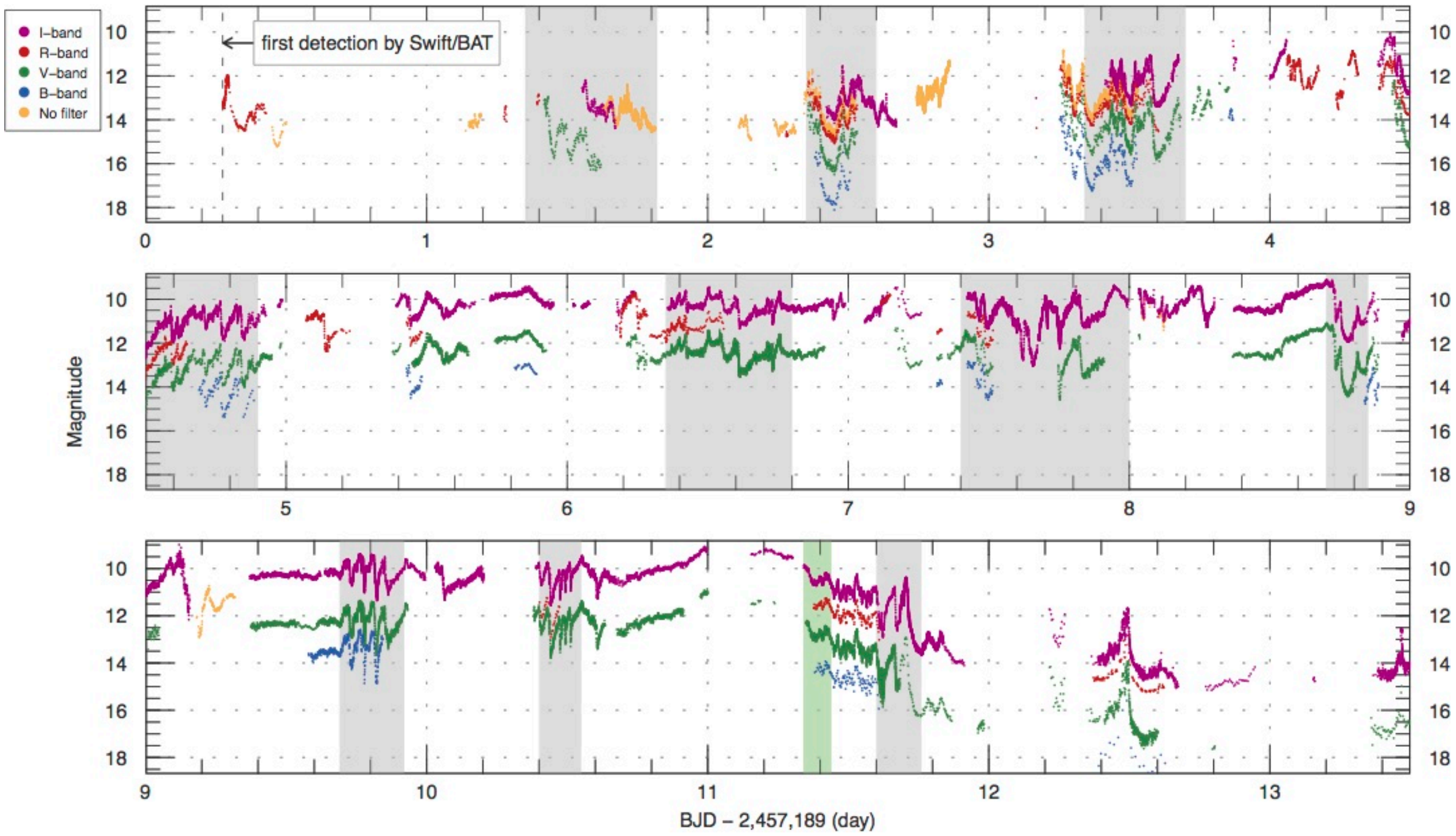

**Figure 1: Overall multi-colour light curves during the 2015 outburst of V404 Cyg.** This figure shows multi-colour light curves (B, V , R, I and no filters) during Barycentric Julian Day (BJD) 2,457,189–2,457,207 (BJD 2,457,189 corresponds to 2015 June 15). We can clearly see that dip-type oscillations (variations with recurrent sudden dips) were observed from beginning to end of the outburst. The horizontal axis represents BJD–2,457,189. The significant periods of repetitive optical variations are indicated in gray and green colours for the "dip-type" and "heartbeat-type" oscillations, respectively.

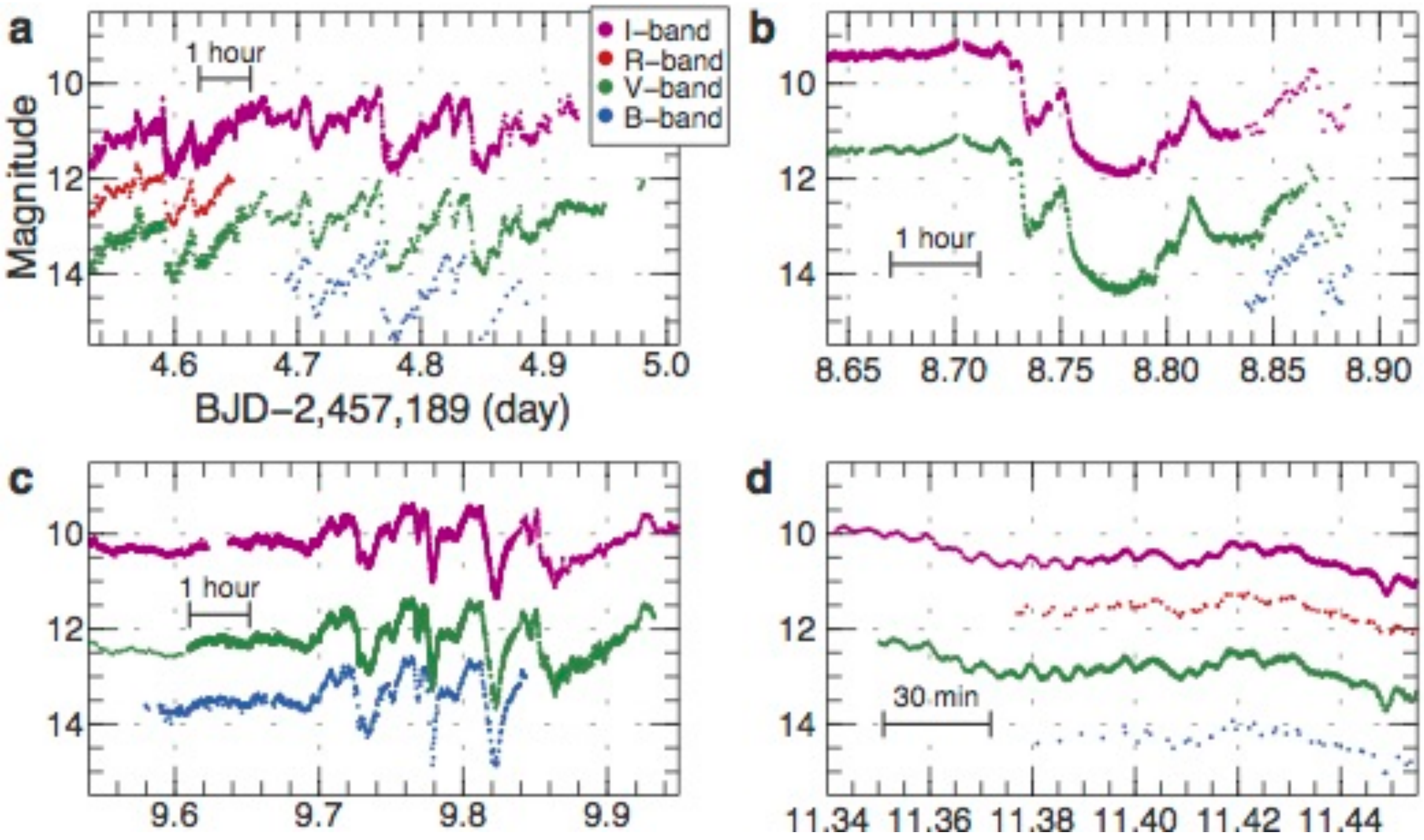

**Figure 2: Short-term and large-amplitude optical variations having repeating structures in the 2015 outburst of V404 Cyg.** Panels a, b, c, and d represent variations with characteristic patterns during BJD 2,457,193.6– 2,457,194.0, 2,457,197.7–2,457,198.0, 2,457,198.6–2,457,198.9, and 2,457,200.34–2,457,200.6, respectively. (a, b, and c) There are gradual rises with increasing amplitudes of fluctuations followed by dips, during which fluctuations disappear. These variations are sometimes accompanied with spikes. The interval between two dips is ∼45 min–∼2.5 hours. (d) Repetitive small oscillations with high coherence are seen at intervals of ∼5 min. The shapes of these oscillations resemble those of GRS 1915+105[3].

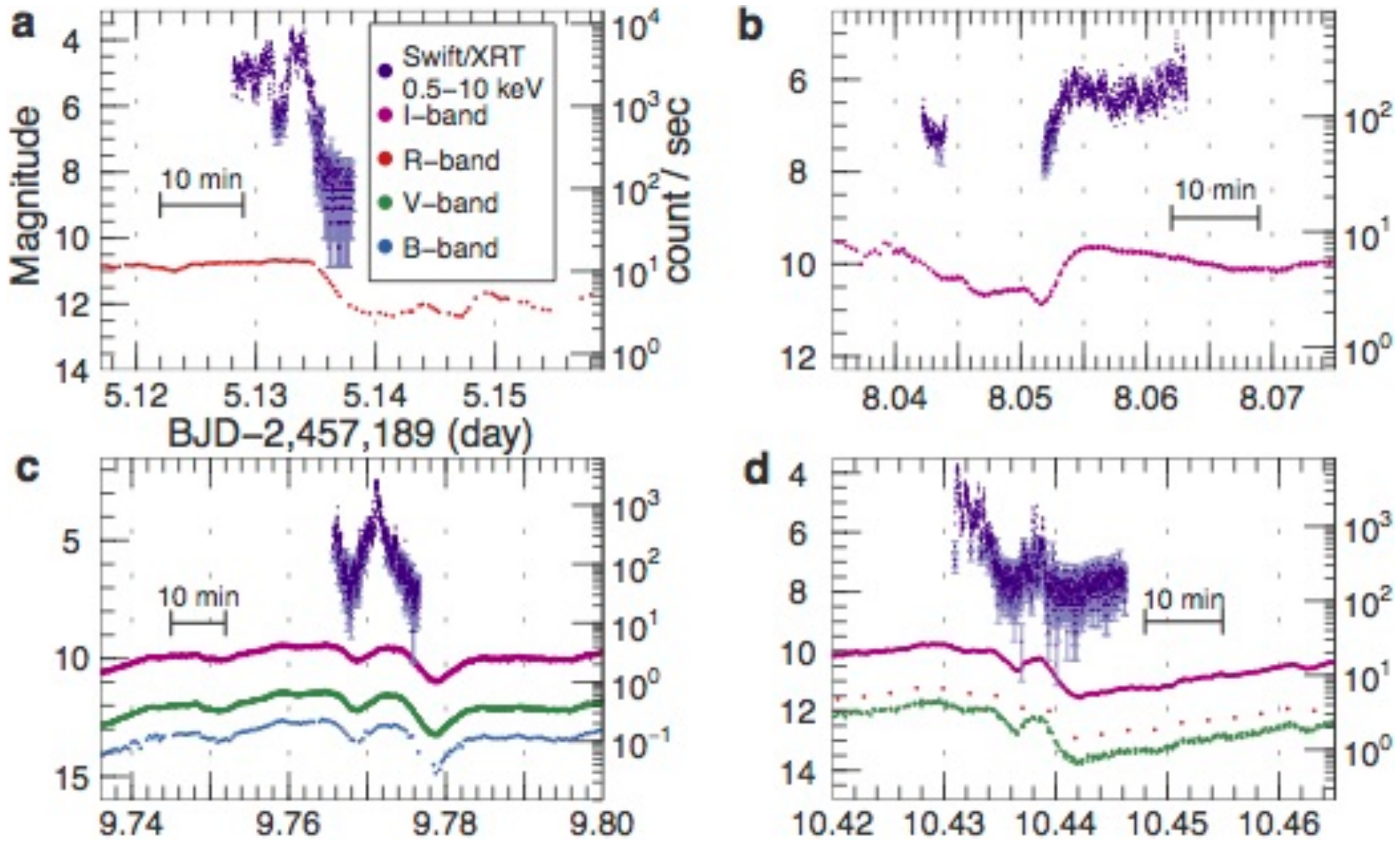

**Figure 3: Correlation between optical and X-ray fluctuations of V404 Cyg in the 2015 outburst.** The terms are (a) BJD 2,457,194.126–2,457,194.140, (b) 2,457,197.050–2,457,197.065, (c) 2,457,198.760–2,457,198.780, and (d) 2,457,199.430–2,457,199.450, respectively. Panels a and b cover the fading and rising phases, respectively. Panels c and d show the correlations of short-term fluctuations. When both X-ray and optical light strongly varied, the correlation is generally good (though note in panels a, c, and d that optical dips lag behind X-ray dips). The navy blue error bars represent 1σ statistics errors. We plot points without errors when errors are sufficiently small.

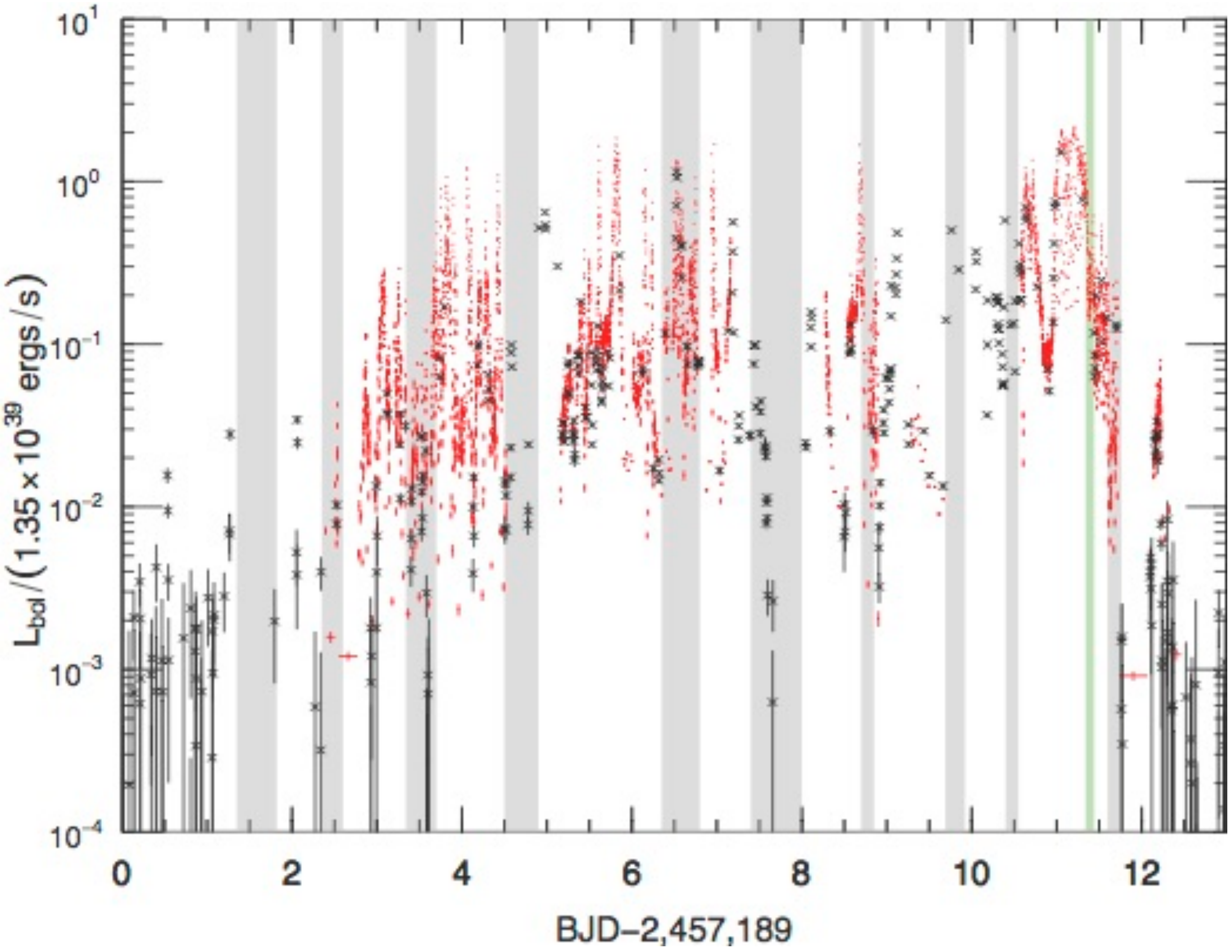

**Figure 4: The bolometric luminosity $L_{bol}$ of V404 Cyg during the 2015 outburst.** It is normalized at the Eddington luminosity assuming a black hole mass of $9M_{\odot}$. The Swift/BAT survey data (15–50 keV) and INTEGRAL Imager on Board the Integral Satellite (IBIS)/CdTe array (ISGRI) monitoring (25–60 keV) are shown in black and red points, respectively. The gray and green markers represent the periods of the "dip-type oscillations" and "heartbeat-type oscillations", respectively. The black error bars represent 1σ statistical errors.

## Methods

## 1. Detailed Methods of Optical Observations

Immediately after the detection by Swift/BAT on June 15.77197 UT, the VSNET collaboration team[31] started a worldwide photometric campaign of V404 Cyg. There was also an independent detection by CCD photometry on June 16.169 UT[32]. Time-resolved CCD photometry was carried out at 27 sites using 36 telescopes with apertures of dozens of centimetres (Extended Data Table 2). We also used the public AAVSO data[33]. We corrected for bias and flat-fielding in the usual manner, and performed standard aperture photometry. The observers except for TAOS[34] used standard filters (B, V, $R_C$, $I_C$; we write R and I for $R_C$ and $I_C$ in the main text and figures for brevity. (Extended Data Table 1) and measured magnitudes of V404 Cyg relative to local comparison stars whose magnitudes were measured by A. Henden (sequence 15167RN) from the AAVSO Variable Star Database[35]. We applied small zero-point corrections to some observers' measurements. When filtered observations were unavailable, we used unfiltered data to construct the light curve. The exposure times were mostly 2–30 s, with some exceptional cases of 120 s in B-band, giving typical time resolutions of a few seconds. All of the observation times were converted to Barycentric Julian Day.

## 2. Comparison with X-ray Observations

For the Swift/(XRT) light curves (Fig. 3 and Extended Data Fig. 2), we extracted source events from a region with a 30-pixel radius centered on V404 Cyg. To avoid pile-up effects, we further excluded an inner circular region if the maximum count rate of the XRT raw light curves, binned in 10 s intervals, exceeded 200 cts $s^{-1}$. The inner radii are set to be 10 and 20 pixels at the maximum raw rate of 1000 cts $s^{-1}$ and 2000 cts $s^{-1}$, respectively, and those for intermediate count rates were determined via linear interpolation between the two points. The presented light curves were corrected for the photon losses due to this exclusion by using the "xrtlccorr" tool. In addition, from panels a, c, and d of Fig. 3, we can see a time delay in the start of a dip in optical light, relative to that in X-rays. The delay time was ~1 min, which is similar to the reported value of 0–50 s[36]. This was determined by cross-correlating the U-band and X-ray (0.3–10 keV) light curves obtained with Swift/UltraViolet and Optical Telescope (UVOT) and Swift/XRT on UT 2015 June 21[36]. The observations were carried out when the source showed little rapid optical flickering and no extreme flares, and thus the nature of the lag may be different from that in our observations. We also note that the apparent difference between the Swift/UVOT and the ground-based times[36] is caused by the drift of the clock on board on the satellite, to which we have applied the necessary corrections.

## 3. Origin of Cyclic Dips: X-ray Spectra Obtained with Swift/XRT during the Optical Observations

In order to examine the possibility that absorption by gas in the line-of-sight causes the observed violent flux variations in the optical and X-ray bands (Fig. 3), we studied intensity-sliced X-ray spectra. As a striking example, we show, in Extended Data Fig. 2a. This period corresponds to that in Fig. 3a when both the X-ray and optical fluxes exhibited a sudden intensity drop toward the latter part of the period. We divided it into five intervals (T1 to T5) (Extended Data Fig. 3a), and generated spectra through the tools "xrtpipeline" and "xrtproducts" for standard pipeline processing. We excluded the central 60-arcsecond strip from this Windowed Timing (WT) mode data, to avoid the heavy pile-up effect when the raw count rate exceeds $\sim 150 \mathrm{ctss}^{-1}$. We compared the $\nu F_\nu$ spectra of the five intervals, where the spectra are fitted by a single power-law model ("pegpwrlw") multiplied by photoelectric absorption ("phabs"). The absorbed X-ray flux ranges by two orders of magnitude from $2.1 \times 10^{-9}$ erg $s^{-1}$ $cm^{-2}$ in T5 to $3.0 \times 10^{-7}$ ergs $s^{-1}$ $cm^{-2}$ in T3. However, the best-fit column density and photon index were relatively stable over the five intervals, st $\sim 2-6 \times 10^{-21}$ $cm^{-2}$ and $\sim$ 1.0–1.5, respectively. Since the X-ray spectrum does not show a noticeable rise in column density when the X-ray flux sharply dropped, and since there is no stronger iron edge in the latter part of the observation, absorption cannot be the primary cause of the time variation in our datasets that cover the X-ray and optical bands simultaneously.

## 4. Objects Showing Violent Short-term Variations in Outburst

We show the list of X-ray binaries which have shown violent short-term variations either in X-rays or in optical wavelengths (Extended Data Table 3).

IGR J17091−3624 is known as the second BH X-ray binary whose X-ray light curves showed a variety of patterns, resembling those of GRS 1915+105[18]. The variations classified as class $\varrho$ ("heartbeat"), class $\nu$ (similar to class $\varrho$ but with secondary peak after the dips), class $\alpha$ ("rounded-bumps"), class $\beta/\lambda$ (repetitive short-term oscillations after low-quiet period), and class $\mu$ were observed in the 2011 outburst[18].

The Rapid Burster (RB or MXB 1730−335), a LMXB containing a neutron star (NS), was discovered by Small Astronomy Satellite (SAS-3) observations[37]. This object has been recently reported to show cyclic long X-ray bursts with periods of a few seconds resembling class ϱ variations ("heartbeat") in GRS 1915+105[24]. Another type of variations which are similar to class θ variations ("M"-shaped light curves) were also observed[24]. The emission of the Rapid Burster did not reach the Eddington luminosity during these variations[38].

V4641 Sgr was originally discovered as a variable star[39] and was long confused with a different variable star, GM Sgr[40]. The object is famous for its short and bright outburst in 1999, which reached a optical magnitude of at least 8.8 mag[41–44]. V4641 Sgr showed short-term variations in optical wavelengths during the 2002, 2003 and 2004 outbursts[14,45–47]. It was the first case in which short-term and large-amplitude variations in the optical range during an outburst were detected. V4641 Sgr is classified as a LMXB, and has a long orbital period. Its mass accretion rate is less than the Eddington rate (except for the 1999 outburst[44,48]). These properties are similar to those of V404 Cyg. However, while the short-term variations of V4641 Sgr seemed to be random, those of V404 Cyg showed repetitive patterns; this is the greatest difference between these two objects. There has been a suggestion that V4641 Sgr is a "microblazar"[49] because the jets observed during the outburst in 1999 were proposed to have the largest bulk Lorentz factor among known galactic sources[43] .

There are also other X-ray transients showing short-term optical variations (e.g., XTE J1118+480 and GX 339−4). However, these two sources are Quasi-Periodic Oscillations (QPOs), characterized by very short periods. The periods are much shorter than those of repetitive patterns (tensof seconds to a few hours), which we discuss in this letter. Furthermore, the amplitudes of their variations are significantly smaller than those observed in V4641 Sgr[4, 50] in timescales longer than tens of seconds.

## 5. Estimation of the Disc Mass and Comparison with the Previous Outbursts

Following the method in [15], we estimated the mass stored in the disc at the onset of the outburst. By integrating the X-ray light curve of Swift/BAT and assuming the spectral model C in Table 1 in [15], we obtained $5.0 \times 10^{25}$ [g] assuming a radiative efficiency of 10 per cent and a distance of 2.4 (±0.2) kpc[8]. The mass during the 1989 outburst has been updated to $3.0 \times 10^{25}$ [g] by using this updated distance. The stored mass in the 2015 outburst was approximately the same as that in the 1989 one. As discussed in [15], these masses are far smaller than the mass of a fully built-up disc, estimated to be $2.0 \times 10^{28}$ [g], if these outbursts were starting at the outermost region.

We compare the published optical light curves of the 1989 and 1938 outbursts[51, 52] with our data from the 2015 outburst (Extended Data Fig. 4). We can see that these outbursts have different durations. The 1938 outburst was apparently longer than the others, and it may have had different properties from the 1989 and 2015 ones. The fading rates of the 1989 and 2015 outbursts are significantly larger than those of classical X-ray transients[6], or FRED-type (fast rise and exponential decline) outbursts, such as 0.028 mag day$^{-1}$ in V518 Per = GRO J0422+32[53] and 0.015 mag day$^{-1}$ in V616 Mon = A0620–00[54]. This supports the hypothesis that the outbursts in 1989 and 2015 are different from typical outbursts of classical X-ray transients and that the stored disc mass was by a factor of $\sim 10^3$ smaller in the 1989 and 2015 outbursts than the mass of a fully built up disc.

## 6. Power Spectrum

We performed power spectral analyses on BJD 2,457,193, BJD 2,457,196, and BJD 2,457,200. We used the continuous and regularly sampled high-cadence dataset obtained by LCO (Extended Data Table 1) with exposure times of 5 s (on BJD 2,457,193) and 2 s (others). The durations of these observations are 1.4, 3.1 and 2.2 hours, respectively. Considering the read-out times of 1 s, the Nyquist frequencies of these observations are 0.08 and 0.17 Hz, respectively. The power spectral densities (PSDs) were calculated using "powspec" software in the FTOOLS Xronos package on magnitude measurements. We did not apply de-trending of the light curve since the durations of the individual observations were shorter than the timescale of the global variation of the outburst. The power spectra are well expressed by a power law [$P \propto f^{-\Gamma}$] with an index $\Gamma$ of 1.9(± 0.1), 1.8(± 0.1), and 2.3(± 0.1) on BJD 2,457,193, 2,457,196, and 2,457,200, respectively (Extended Data Fig. 5). Interpretation of the physical origins based of these variations is difficult because a power law index of ≈2 in the PSDs is often observed in natural phenomena. In this region ($f < 0.01$ Hz), the power originating in the optical variations of V404 Cyg is significantly higher than that of white noise estimated from the observations.

We next summarize the other reports on short-term variations of V404 Cyg during the present outburst. On BJD 2,457,191, this object was observed using the Argos photometer on the 2.1m Otto Struve Telescope at McDonald Observatory with an exposure time of 2 s[55]. They reported that the power spectrum was dominated by steep red noise. Observations on BJD 2,457,193 and BJD 2,457,194 were also performed using the ULTRACAM attached with the 4.2m William Herschel Telescope on La Palma observatory with a high time resolution (466.8 ms)[56]. They reported that the variations were dominated by timescales longer than tens of seconds. Although large amplitude flares (0.3–0.4 mag) on time scales shorter than 1 s were reported[57], these flares may be of different origin. For the variations with timescales longer than 100 s, our results agree with these reports[55, 56].

## 7. Disc Radius Inferred from Final Fading Rate

The timescale $\tau$ of heating/cooling waves in dwarf novae and X-ray transients[58] is a function of the central mass ($M_1$) and radius ($r$) with the form, $\tau \propto \alpha M^{-1/2} r^{3/2}$, where $\alpha$ is the viscosity parameter[59]. Here, we estimate the disc radius of V404 Cyg assuming that the timescale of the final fading reflected a dwarf nova-type cooling wave. Using the Kepler data of V344 Lyr[60] and V1504 Cyg, we measured a fading rate of 1.5 mag day$^{-1}$ of the normal outbursts immediately preceding superoutbursts. During these outbursts, the disc radius is expected to be very close to the 3:1 resonance radius. Adopting a typical mass of a white dwarf in a cataclysmic variable ($M_1 = 0.83 M_\odot$[61]), we estimated the disc radius of V404 Cyg to be $7.8 \times 10^{10}$ cm for a black hole mass of 9.0 $M_\odot$. This size is much smaller than the radius $1.2 \times 10^{12}$ cm which is expected for a fully built-up disc[15].

## 8. Spectral Energy Distribution Modeling

Extended Data Fig. 6a shows the multi-wavelength spectral energy distribution on BJD 2,457,199.431– 2,457,199.446, when the source was simultaneously observed in the X-ray, ultraviolet (U V ), and optical bands. The optical fluxes in the V and $I_C$ bands are taken from our photometric data averaged over the period. Note that $R_C$-band data are also available but not used here, because of the contamination of the continuum strong Hα line[62–64].

The X-ray spectrum is extracted from simultaneous Swift/XRT data (ObsID 00031403058), which were taken in the Windowed Timing mode. The data are processed through the pipeline processing tool "xrtpipeline". The events detected within 20 pixels around the source position are removed to mitigate pileup effects. The U -band flux is obtained from the Swift/UVOT images with the same ObsID as the XRT, through the standard tool "uvot2pha" provided by the Swift team. A circular region centred at the source position with a radius of 5 arcsec is adopted as the source extraction region of the UVOT data. The optical, UV, and X-ray data are corrected for interstellar extinction/absorption by assuming $A_V = 4$[65] and using the extinction curve in [66] and the $N_H$ versus $E(B - V)$ relation in [67]. Radio data are from the RATAN-600 observation performed in the same period[68] .

The multi-wavelength SED can be reproduced with the "diskir" model[69,70], which accounts for the emission from the accretion disc, including the effects of Comptonisation in the inner disc and reprocessing in the outer disc. We find that partial covering X-ray absorption (using the "pcfabs" model implemented in the spectral analysis software XSPEC) improves the quality of the fit significantly. The inner disc temperature is estimated to be 0.12 ± 0.01 keV, and the electron temperature and photon index of the Comptonisation component, the ratio between the luminosity of the Compton tail and disc blackbody ($L_C/L_d$), and the fraction of the bolometric flux thermalized in the outer disc ($f$), are 17.5±0.8 keV, 1.78±0.03, 1.17±0.03, and $1.3^{+0.6}_{-0.8} \times 10^{-2}$, respectively (the errors in this section represent 90% confidence ranges for one parameter). The inner radius ($R_{in}$) is estimated to be $(1.5 - 5.4) \times 10^8$ cm, and the outer radius ($R_{out}$) is $(2.5 \pm 0.3) \times 10^{12}$ cm. The derived value of $R_{out}$ is comparable to or even larger than the binary separation ($\sim 2.2 \times 10^{12}$ cm). However, it can be smaller due to uncertainties in interstellar/circumbinary extinction[71] and/or the contribution of jet emission. For instance, if $A_V$ is 0.4 magnitude larger than the assumed value (4.0), $R_{out}$ becomes $1.9 \pm 0.2 \times 10^{12}$ cm. The maximum achievable radius of a stable disc for a $q$ (mass ratio) = 0.06 object (Extended Data Table 3) is around 0.62A (radius of the 2:1 resonance) to ~0.7A (tidal limit), where A is the binary separation[72]. Considering the uncertainties, the result of our analysis ($\gtrsim$ 0.77A) is compatible with this maximum radius. Our result appears to favour a large $A_V$ value. For the partial covering absorber, the best-fit value of the column density is $5.2^{+0.4} \times 10^{23}$ cm$^{-2}$ and that of the covering fraction is 64 ± 4%.

The radio SED can be approximated by a power-law with a photon index of ≈1, as in other black hole binaries in the low/hard state[73]. This profile is likely generated by the optically-thick synchrotron emission from compact jets[74]. Because optically-thick synchrotron spectrum often extends up to the mm to near-infrared bands[75–77], it may contribute to the optical fluxes, in particular at longer wavelengths. The blackbody emission from the companion, a K3III-type star[7] with a radius of $\sim 3\ R_\odot$ and a temperature of ~4320 K, contributes to the SED only negligibly.

Extended Data Fig. 6b plots the simultaneous SED on BJD 2,457,191.519–2,457,191.524, which is ~ 2 orders of magnitude fainter in the X-ray band than that shown in the left panel. The X-ray, UV, and optical data are taken from the Swift data (ObsID 00031403038) and our photometric measurements in same manner as described above. This SED can be reproduced with the irradiated disc model as well, with somewhat smaller photon index ($1.43^{+0.02}_{-0.03}$) and inner disc temperature (< 0.07 keV), and a larger f ($0.06^{+0.02}_{-0.05}$) than those on BJD 2,457,199.431–2,457,199.446.

## 9. Time History of the Bolometric Luminosity

The bolometric luminosity $L_{bol}$ of V404 Cyg is evaluated based on the hard X-rays above ~15 keV where the intrinsic spectrum is less affected by an absorption. We processed the Swift/BAT archival survey data via "batsurvey" in the HEAsoft package to derive count rates with individual exposures of ~300 seconds. Even within this short exposure, photon statistics are good during bright states (>0.05 counts $s^{-1}$). Assuming a Crab-like spectrum (1 Crab~ 0.039 counts $s^{-1}$), the BAT count rates R (counts $s^{-1}$) are then converted into 15–50keVflux($F_{15-50}$) and luminosity($L_{15-50}$) using $F_{15-50} = 3.6\times10^{-7} R (erg s^{-1} cm^{-1})$and a fiducial distance of 2.4 kpc, respectively. In Fig. 4, we show $L_{bol}$ after multiplying by a conversion factor $L_{bol}/L_{15-50} = 7$ determined from SED modeling (Sec. 8 in Methods). We find that this bolometric correction factor lies within the range of 2.5–10 by fitting nineteen X-ray(XRT)-optical simultaneous SED in different periods between BJD 2,457,192.019 and 2,457,201.011. Since the BAT survey data are rather sparse, in order to catch shorter-term variations, we further overlaid the INTEGRAL IBIS/ISGRI monitoring in the 25–60 keV band available at[78], assuming a conversion parameter of 1 Crab rate to be 172.1 counts $s^{-1}$ and a bolometric correction factor at 9.97.

The luminosity was highly variable during the outburst, changing by five orders of magnitude. While V404 Cyg sometimes reaches the Eddington luminosity ($L_{Edd}$) at the peak of multiple sporadic flares, it also repeatedly dropped below 1–10% of $L_{Edd}$ (Fig. 4). At earlier phases of this outburst, the characteristic oscillation already occurred during a lower luminosity state as discussed in the main text.

## Extended Data tables:

**Extended Data Table 1: A log of photometric observations of the 2015 outburst of V404 Cyg.** Start and end dates of observations, mean magnitudes, 1σ of mean magnitudes, numbers of observations, observers' codes, and filters are summarized. Note that observers for TAOS used custom made filters close to the union of standard R and V [34], but the magnitude reported in this paper was approximately calibrated to standard R.

| Start* | End* | Mag† | Error‡ | N§ | Obs‖ | Band¶ |
|---|---|---|---|---|---|---|
| 0.274 | 0.295 | 13.24 | 0.032 | 215 | TAO | R |
| 0.282 | 0.499 | 15.34 | 0.101 | 37 | PZN | CR |
| 0.386 | 0.426 | 15.18 | 0.036 | 86 | PZN | $R_C$ |
| 0.386 | 0.426 | 14.31 | 0.040 | 20 | CRI&PZN | $R_C$ |
| 1.137 | 1.192 | 14.92 | 0.024 | 61 | PZN | CR |
| 1.274 | 1.398 | 13.93 | 0.086 | 20 | PZN | $R_C$ |
| 1.551 | 1.670 | 13.48 | 0.029 | 191 | deM | $I_C$ |
| 1.627 | 1.810 | 14.98 | 0.009 | 2430 | LCO | CR |
| 2.109 | 2.517 | 15.07 | 0.024 | 224 | PZN | CR |
| 2.277 | 2.404 | 14.57 | 0.104 | 35 | PZN | $R_C$ |
| 2.341 | 2.522 | 14.19 | 0.044 | 231 | DPV | CR |
| 2.354 | 2.529 | 15.41 | 0.056 | 158 | DPV | V |
| 2.354 | 2.529 | 14.12 | 0.049 | 158 | DPV | $R_C$ |
| 2.380 | 2.505 | 17.19 | 0.075 | 61 | Ter | B |
| 2.380 | 2.505 | 15.52 | 0.072 | 61 | Ter | V |
| 2.381 | 2.506 | 14.33 | 0.062 | 61 | Ter | $R_C$ |
| 2.406 | 2.524 | 14.65 | 0.024 | 354 | Ter | CR |
| 2.422 | 2.615 | 14.43 | 0.045 | 151 | Kai | $I_C$ |
| 2.423 | 2.609 | 14.43 | 0.045 | 147 | Kai | $R_C$ |
| 2.446 | 2.669 | 13.46 | 0.021 | 667 | deM | $I_C$ |
| 2.742 | 2.859 | 13.91 | 0.009 | 2652 | LCO | CR |
| 3.801 | 3.341 | 12.55 | 0.048 | 1216 | TAO | R |
| 3.251 | 3.524 | 16.10 | 0.054 | 186 | Ter | B |
| 3.252 | 3.525 | 14.45 | 0.051 | 183 | Ter | V |
| 3.252 | 3.524 | 13.41 | 0.044 | 177 | Ter | $R_C$ |
| 3.260 | 3.529 | 13.58 | 0.017 | 1278 | Ter | CR |
| 3.266 | 3.308 | 13.54 | 0.086 | 48 | PZN | CR |
| 3.271 | 3.307 | 13.64 | 0.091 | 40 | PZN | $R_C$ |
| 3.410 | 3.489 | 15.80 | 0.095 | 38 | CRI | B |
| 3.411 | 3.488 | 14.36 | 0.071 | 37 | CRI | V |
| 3.411 | 3.488 | 13.17 | 0.062 | 37 | CRI | $R_C$ |
| 3.411 | 3.489 | 12.01 | 0.058 | 37 | CRI | $I_C$ |
| 3.419 | 3.588 | 14.48 | 0.048 | 189 | RPc | V |
| 3.428 | 3.553 | 14.52 | 0.056 | 128 | Trt | V |
| 3.430 | 3.519 | 12.25 | 0.023 | 597 | IMi | $I_C$ |
| 3.435 | 3.673 | 12.47 | 0.020 | 1036 | deM | $I_C$ |
| 3.525 | 3.650 | 12.64 | 0.075 | 37 | COO | $I_C$ |
| 3.530 | 3.820 | 14.53 | 0.076 | 165 | Kis | V |
| 3.819 | 3.821 | 10.39 | 0.014 | 2 | Kis | $I_C$ |
| 3.998 | 4.057 | 11.49 | 0.038 | 149 | KU1 | $I_C$ |
| 4.059 | 4.311 | 12.04 | 0.036 | 397 | Mdy | $R_C$ |
| 4.187 | 4.316 | 11.88 | 0.022 | 169 | TAO | R |
| 4.435 | 4.673 | 11.66 | 0.021 | 1089 | deM | $I_C$ |
| 4.546 | 4.649 | 13.41 | 0.041 | 82 | Kis | V |
| 4.579 | 4.637 | 11.32 | 0.008 | 1416 | LCO | $I_C$ |
| 4.976 | 4.978 | 12.14 | 0.034 | 5 | Kis | V |
| 4.979 | 4.981 | 10.04 | 0.004 | 3 | Kis | $I_C$ |
| 5.070 | 5.223 | 11.19 | 0.042 | 254 | Mdy | $R_C$ |
| 5.426 | 5.481 | 12.75 | 0.054 | 36 | CRI | B |
| 5.427 | 5.481 | 13.96 | 0.057 | 36 | CRI | V |
| 5.427 | 5.480 | 11.66 | 0.048 | 35 | CRI | $R_C$ |
| 5.427 | 5.480 | 10.57 | 0.044 | 36 | CRI | $I_C$ |
| 5.448 | 5.633 | 10.45 | 0.010 | 840 | deM | $I_C$ |
| 5.595 | 5.670 | 10.25 | 0.020 | 25 | COO | $I_C$ |
| 5.724 | 5.954 | 9.85 | 0.007 | 920 | SWI | $I_C$ |
| 5.745 | 5.911 | 11.69 | 0.011 | 346 | GFB | V |
| 5.923 | 5.949 | 10.32 | 0.013 | 24 | COO | $I_C$ |
| 6.011 | 6.015 | 12.51 | 0.016 | 5 | OKU | V |
| 6.019 | 6.076 | 10.27 | 0.005 | 154 | OKU | V |
| 6.146 | 6.157 | 10.01 | 0.050 | 4 | KW2 | $I_C$ |
| 6.146 | 6.157 | 12.02 | 0.121 | 4 | KW2 | V |
| 6.146 | 6.157 | 13.15 | 0.159 | 2 | KW2 | B |
| 6.182 | 6.281 | 10.55 | 0.048 | 129 | Aka | $R_C$ |
| 6.210 | 6.280 | 12.41 | 0.060 | 64 | Aka | V |
| 6.293 | 6.554 | 11.31 | 0.030 | 85 | CRI | $R_C$ |
| 6.295 | 6.550 | 12.30 | 0.037 | 83 | CRI | V |
| 6.346 | 6.428 | 11.84 | 0.028 | 93 | PZN | $R_C$ |
| 3.356 | 6.543 | 9.94 | 0.011 | 412 | DPV | $I_C$ |
| 6.363 | 6.521 | 12.25 | 0.010 | 572 | Trt | V |
| 6.369 | 6.406 | 10.09 | 0.008 | 334 | DPV | $I_C$ |
| 6.430 | 6.615 | 12.38 | 0.022 | 418 | RJV | V |
| 6.584 | 6.827 | 12.65 | 0.005 | 5910 | LCO | V |
| 6.592 | 6.861 | 10.54 | 0.012 | 794 | RIT | $I_C$ |
| 6.717 | 6.944 | 10.36 | 0.007 | 942 | SWI | $I_C$ |
| 6.745 | 6.912 | 12.35 | 0.010 | 347 | GFB | V |
| 6.919 | 6.950 | 10.35 | 0.076 | 24 | COO | $I_C$ |
| 7.056 | 7.057 | 13.27 | 0.010 | 3 | Kis | V |
| 7.057 | 7.137 | 10.55 | 0.019 | 295 | Kis | V |
| 7.115 | 7.147 | 10.01 | 0.018 | 45 | Aka | $R_C$ |
| 7.144 | 7.150 | 9.87 | 0.016 | 18 | KW2 | $I_C$ |
| 7.144 | 7.150 | 11.76 | 0.068 | 18 | KW2 | V |
| 7.144 | 7.150 | 13.44 | 0.184 | 2 | KW2 | B |
| 7.313 | 7.512 | 13.79 | 0.063 | 67 | CRI | B |
| 7.314 | 7.511 | 12.54 | 0.065 | 67 | CRI | V |
| 7.314 | 7.512 | 11.45 | 0.060 | 66 | CRI | $R_C$ |
| 7.315 | 7.511 | 10.36 | 0.061 | 65 | CRI | $I_C$ |
| 7.422 | 7.588 | 10.52 | 0.011 | 1501 | IMi | $I_C$ |
| 7.427 | 7.670 | 11.35 | 0.035 | 1104 | deM | $I_C$ |
| 7.675 | 7.802 | 10.81 | 0.012 | 1961 | LCO | $I_C$ |
| 7.707 | 7.945 | 10.70 | 0.019 | 1003 | SWI | $I_C$ |
| 7.744 | 7.907 | 12.98 | 0.036 | 350 | GFB | V |
| 8.030 | 8.300 | 10.22 | 0.020 | 535 | KU1 | $I_C$ |
| 8.032 | 8.035 | 11.69 | 0.036 | 5 | OKU | V |
| 8.038 | 8.297 | 10.27 | 0.016 | 1022 | OKU | $I_C$ |
| 8.038 | 8.128 | 12.03 | 0.040 | 81 | Ioh | CR |
| 8.152 | 8.214 | 12.41 | 0.028 | 103 | Wnm | cG |
| 8.360 | 8.543 | 9.95 | 0.015 | 68 | CRI | $I_C$ |
| 8.394 | 8.619 | 10.41 | 0.011 | 623 | Kai | $I_C$ |
| 8.419 | 8.671 | 10.17 | 0.012 | 1129 | deM | $I_C$ |
| 8.709 | 8.859 | 13.42 | 0.036 | 413 | RIT | V |
| 8.969 | 9.043 | 10.87 | 0.019 | 296 | Sac | $I_C$ |
| 8.993 | 9.154 | 10.55 | 0.024 | 608 | Kis | $I_C$ |
| 9.006 | 9.044 | 12.77 | 0.032 | 40 | Sac | V |
| 9.179 | 9.315 | 12.49 | 0.053 | 146 | PZN | CR |
| 9.224 | 9.229 | 12.59 | 0.149 | 5 | OKU | V |
| 9.239 | 9.300 | 10.84 | 0.020 | 152 | OKU | $I_C$ |
| 9.382 | 9.620 | 10.40 | 0.002 | 643 | Kai | $I_C$ |
| 9.414 | 9.595 | 10.24 | 0.003 | 428 | NDJ | $I_C$ |
| 9.577 | 9.841 | 13.54 | 0.020 | 620 | RIT | B |
| 9.607 | 9.798 | 12.21 | 0.005 | 4709 | LCO | V |
| 9.635 | 9.828 | 10.13 | 0.010 | 1823 | LCO | $I_C$ |
| 9.744 | 9.911 | 12.38 | 0.031 | 350 | GFB | V |
| 10.027 | 10.028 | 11.90 | 0.018 | 3 | Kis | V |
| 10.029 | 10.201 | 10.54 | 0.011 | 837 | Kis | $I_C$ |
| 10.387 | 10.619 | 10.46 | 0.020 | 611 | Kai | $I_C$ |
| 10.415 | 10.670 | 10.38 | 0.013 | 1389 | deM | $I_C$ |
| 10.744 | 10.910 | 11.99 | 0.010 | 349 | GFB | V |
| 11.182 | 11.300 | 9.41 | 0.012 | 99 | KU1 | $I_C$ |
| 11.291 | 11.298 | 10.55 | 0.003 | 112 | TAO | R |
| 11.339 | 11.514 | 10.51 | 0.018 | 406 | DPV | $I_C$ |
| 11.348 | 11.554 | 13.10 | 0.014 | 730 | Trt | V |
| 11.372 | 11.515 | 13.15 | 0.019 | 335 | DPV | V |
| 11.385 | 11.592 | 11.00 | 0.015 | 490 | Kai | $I_C$ |
| 11.421 | 11.673 | 11.32 | 0.021 | 1314 | deM | $I_C$ |
| 11.460 | 11.624 | 13.71 | 0.097 | 70 | JSa | V |
| 11.483 | 11.603 | 13.53 | 0.016 | 374 | RJV | V |
| 11.590 | 11.679 | 14.43 | 0.014 | 730 | LCO | V |
| 11.679 | 11.834 | 12.95 | 0.008 | 3859 | LCO | $I_C$ |
| 12.228 | 12.232 | 15.49 | 0.139 | 5 | OKU | V |
| 12.234 | 12.271 | 12.95 | 0.028 | 177 | TAO | R |
| 12.302 | 12.334 | 13.81 | 0.011 | 311 | TAO | R |
| 12.386 | 12.611 | 13.89 | 0.025 | 484 | Kai | $I_C$ |
| 12.405 | 12.670 | 14.08 | 0.022 | 640 | deM | $I_C$ |
| 12.484 | 12.599 | 16.87 | 0.031 | 237 | RJV | V |
| 13.058 | 13.314 | 15.81 | 0.013 | 211 | Mdy | $R_C$ |
| 13.199 | 13.334 | 15.97 | 0.048 | 1772 | TAO | R |
| 13.382 | 13.594 | 14.28 | 0.014 | 467 | Kai | $I_C$ |
| 13.415 | 13.670 | 14.08 | 0.022 | 640 | deM | $I_C$ |
| 13.438 | 13.473 | 13.91 | 0.040 | 93 | NDJ | $I_C$ |
| 14.014 | 14.021 | 17.21 | 0.075 | 5 | OKU | V |
| 14.026 | 14.168 | 14.75 | 0.022 | 97 | OKU | $I_C$ |
| 14.043 | 14.276 | 14.70 | 0.016 | 361 | Kis | $I_C$ |
| 14.379 | 14.499 | 17.00 | 0.030 | 52 | Trt | CV |
| 14.421 | 14.565 | 14.85 | 0.012 | 152 | RPc | $I_C$ |
| 14.422 | 14.614 | 14.56 | 0.007 | 248 | NDJ | $I_C$ |
| 14.504 | 14.517 | 17.15 | 0.114 | 5 | Trt | V |
| 14.601 | 14.810 | 15.56 | 0.003 | 1830 | LCO | CR |
| 15.166 | 15.276 | 16.92 | 0.234 | 664 | TAO | R |
| 15.356 | 15.549 | 16.04 | 0.007 | 244 | DPV | CR |
| 15.364 | 15.550 | 17.41 | 0.024 | 42 | DPV | V |
| 15.434 | 15.559 | 14.84 | 0.017 | 81 | RPc | $I_C$ |
| 15.694 | 15.762 | 14.45 | 0.008 | 166 | SWI | $I_C$ |
| 16.092 | 16.142 | 14.60 | 0.397 | 5 | TAO | R |
| 16.302 | 16.377 | 16.07 | 0.013 | 26 | PZN | CR |
| 16.320 | 16.525 | 14.44 | 0.010 | 129 | CRI | $I_C$ |
| 16.344 | 16.435 | 14.50 | 0.012 | 52 | DPV | $I_C$ |
| 16.516 | 16.530 | 14.44 | 0.030 | 6 | RPc | $I_C$ |
| 16.680 | 16.937 | 14.36 | 0.006 | 335 | SWI | $I_C$ |
| 17.358 | 17.518 | 14.48 | 0.006 | 218 | DPV | $I_C$ |
| 17.418 | 17.671 | 14.78 | 0.006 | 309 | deM | $I_C$ |
| 17.440 | 17.575 | 14.62 | 0.014 | 43 | RPc | $I_C$ |
| 18.297 | 18.336 | 17.37 | 0.254 | 470 | TAO | R |
| 19.328 | 19.332 | 16.45 | 0.258 | 68 | TAO | R |
| 19.403 | 19.451 | 14.82 | 0.011 | 33 | DPV | $I_C$ |
| 19.423 | 19.498 | 14.79 | 0.026 | 17 | RPc | $I_C$ |
| 19.712 | 19.761 | 16.61 | 0.008 | 60 | GFB | CV |
| 20.435 | 20.592 | 14.98 | 0.008 | 90 | RPc | $I_C$ |
| 21.023 | 21.031 | 15.34 | 0.012 | 10 | RPc | $I_C$ |

∗ JD−2457189. † Mean magnitude. ‡ 1σ of mean magnitude. § Number of observations.

‖ Observer's code: PZN (IKI GRB follow up network), CRI (Crimean Observatory Team), deM (Enrique de Miguel), DPV (Pavol A. Dubovsky), Ter (Terskol Observatory), Kai (Kiyoshi Kasai), NDJ (Nick James), RPc (Roger D. Pickard), Trt (Tam´as Tordai), COO (Lew Cook), Kis (Seiichiro Kiyota), KU1 (Kyoto U. Team), Mdy (Yutaka Maeda), LCO (Colin Littlefield), RIT (Michael Richmond), RJV (Ruiz F. Javier), GFB (William Goff), SWI (William L. Stein), OKU (Osaka Kyoiku U. team), Sac (Atsushi Miyashita), IMi (Ian Miller), TAO (TAOS Team), KW2 (Hiroyuki Maehara), Aka (Hidehiko Akazawa), Wnm (Makoto Watanabe), Ioh (Hiroshi Itoh) and JSa (J. Lluis)

¶ Filter. *B*, *V* , $R_C$, $I_C$ are the standard Johnson-Cousins system. "CR" and "CV" mean unfiltered CCD photometry adjusted R and V for the zero point, respectively. "cG" means green (G) channel output in digital single-lens reflex camera, which gives an approximate response close to V [78] (Kloppenborg et al. 2012).

| CODE | Telescope (& CCD) | Observatory (or Observer) | Site |
|---|---|---|---|
| PZN | 1m Zeiss-1000 Tien Shan +Apogee Alta | Astronomical Observatory | Almaty, Kazakhstan |
| | 40cm ORI-40+FLI ML09000 | ISON-Khureltogot | Mongolia |
| | 70cm+FLI AS-32+FLI IMG6303E | Abastumani observatory | Georgia |
| CRI | 1.25m AZT-11+FLI ProLine PL230 | Crimean astrophysical observatory | Crimea |
| | 38cm K-380+Apogee E47 | Cremean astrophysical observatory | Crimea |
| deM | 35cm SC+QSI-516wsg | Observatorio Astronomico del CIECEM | Huelva, Spain |
| DPV | 28cm SC+MII G2-1600 | Astronomical Observatory on Kolonica | Slovakia |
| | 35cm SC+MII G2-1600 | Astronomical Observatory on Kolonica | Slovakia |
| | VNT 1m+FLI PL1001E | Astronomical Observatory on Kolonica | Slovakia |
| Ter | Zeiss-600 60cm+SBIG STL-1001E | Terskol Observatory | Russia |
| | S2C 35cm | Terskol Observatory | Russia |
| Kai | 28cm SC+ST7XME | Kiyoshi Kasai | Switzerland |
| NDJ | 28cm SC+ST9XE | Nick James | UK |
| RPc | FTN 2.0m+E2V 42-40 | LCOGT* | Hawaii, USA |
| | 35cmSC+SXV-H9 CCD | Roger D. Pickard | UK |
| Trt | 25cm ALCCD5.2 (QHY6) | Tamás Tordai | Budapest, Hungary |
| COO | T07† 43cm+STL-1100M | AstroCamp Observatory | Nerpio, Spain |
| | T21† 43cm+FLI-PL6303E | iTelescope.Net Mayhill | New Mexico, USA |
| | T11† 50cm+FLI ProLine PL11002M | iTelescope.Net Mayhill | New Mexico, USA |
| Kis | 25cm SC+Alta F47 | Seiichiro Kiyota | Kamagaya, Japan |
| | T18† 32cm+STXL-6303E | AstroCamp Observatory | Nerpio, Spain |
| | T5† 25cm+ST-10XME | iTelescope.Net Mayhill | New Mexico, USA |
| | T24† 61cm+FLI-PL09000 | Sierra Remote Observatoy | California, USA |
| KU1 | 40cm SC+ST-9XEI | Kyoto U. Team | Kyoto, Japan |
| Mdy | 35cm SC+ST10XME | Yutaka Maeda | Nagasaki, Japan |
| LCO | 60cm+Apogee Alta U42 CCD | Van Vleck Observatory | Connecticut, USA |
| | 40cm+SBIG STL-6303 | Van Vleck Observatory | Connecticut, USA |
| RIT | 30cm+ST-9E | RIT Observatory | New York, USA |
| RJV | LX200R 40cm+ST8 XME | Observatorio de Cantabria | Spain |
| GFB | CDK 50cm+Apogee U6 | William Goff | California, USA |
| SWI | C14 35cmSC+ST10XME | William L. Stein | New Mexico, USA |
| OKU | 51cm+Andor DW936N-BV | OKU Astronomical Observatory | Osaka, Japan |
| Sac | 20cmL+ST-7XMEi | Atsushi Miyashita | Tokyo, Japan |
| IMi | 35cm SC+SXVR-H16 | Furzehill Observatory | UK |
| TAO | TAOS-B‡ 50cm+SI800 E2V47-20 | Lulin Observatory | Taiwan |
| | TAOS-D‡ 50cm+SI800 E2V47-20 | Lulin Observatory | Taiwan |

∗ Las Cumbres Observatory Global Telescope Network

† itelescope.net

‡ The Taiwanese-American Occultation Survey (TAOS)[81] (Alcock et al. 2003); [34] (Lehner et al. 2009); [82] (Zhang et al. 2013)

**Extended Data Table 2: List of instruments for optical observations.** Observers' codes (CODE) (see Extended Data Table 1), names of telescopes & CCD cameras, observatory (or observer) and sites are summarized.

| | V404 Cyg | GRS 1915+105 | IGR J17091-3624 | Rapid Burster | V4641 Sgr |
|---|---|---|---|---|---|
| **Orbital period [d]** | 6.47129(7)[1] | 33.85(16)[2] | >4[3] | – | 2.81678[4] |
| **Compact object** | BH | BH | BH | NS | BH |
| **Spectrum of the secondary** | K3III[5] | K–M[2] | – | – | B9III[4] |
| **$M_1$ ( $M_\odot$)** | 9.0(0.6)[5] | 10.1(0.6)[6] | 11.8–13.7[7] | 1.1(0.3)[8] | 7.1(0.3)[9] |
| **$q = M_2 / M_1$ (Mass ratio)** | 0.06[5] | 0.042(0.024)[2] | – | – | 0.45(0.05)[9] |
| **$i$ [deg] (Inclination angle)** | 67(3)[5] | 66(2)[10] | 50–70[11] | – | 72.3(4.1)[9] |
| ***V* magnitude minimum** | 18.4[12] | – | – | – | 13.8[9] |
| ***V* magnitude maximum** | 10.9[13] | – | – | – | 8.8[14] |

**References:** 1. Casares et al. (1994) [83]; 2. Steeghs et al. (2013) [20]; 3. Wijnands et al. (2012) [84]; 4. Orosz et al. (2001) [23]; 5. Khargharia et al. (2010) [7]; 6. Reid et al. (2014) [85]; 7. Iyer et al. (2015) [86]; 8. Sala et al. (2012) [87]; 9. MacDonald et al. (2014) [88]; 10. Fender et al. (1999) [89]; 11. King et al. (2012) [90]; 12. Szkody et al. (1989) [91]; 13. This work; 14. Stubbings et al. (1999) [41]

**NOTE:** $M_1$ and $M_2$ represent mass of the central object and that of the secondary star, respectively.

**Extended Data Table 3: Basic information on objects showing violent short-term variations in outbursts.** Orbital period, nature of the compact object, spectrum of the secondary, mass of the central object ($M_1$), mass ratio ($q$), inclination angle ($i$), minimum magnitude (V-band), and maximum magnitude (V-band) on V404 Cyg, GRS 1915+105, IGR 17091−3624, the Rapid Burster, and V4641 Sgr.

## Extended Data figures:

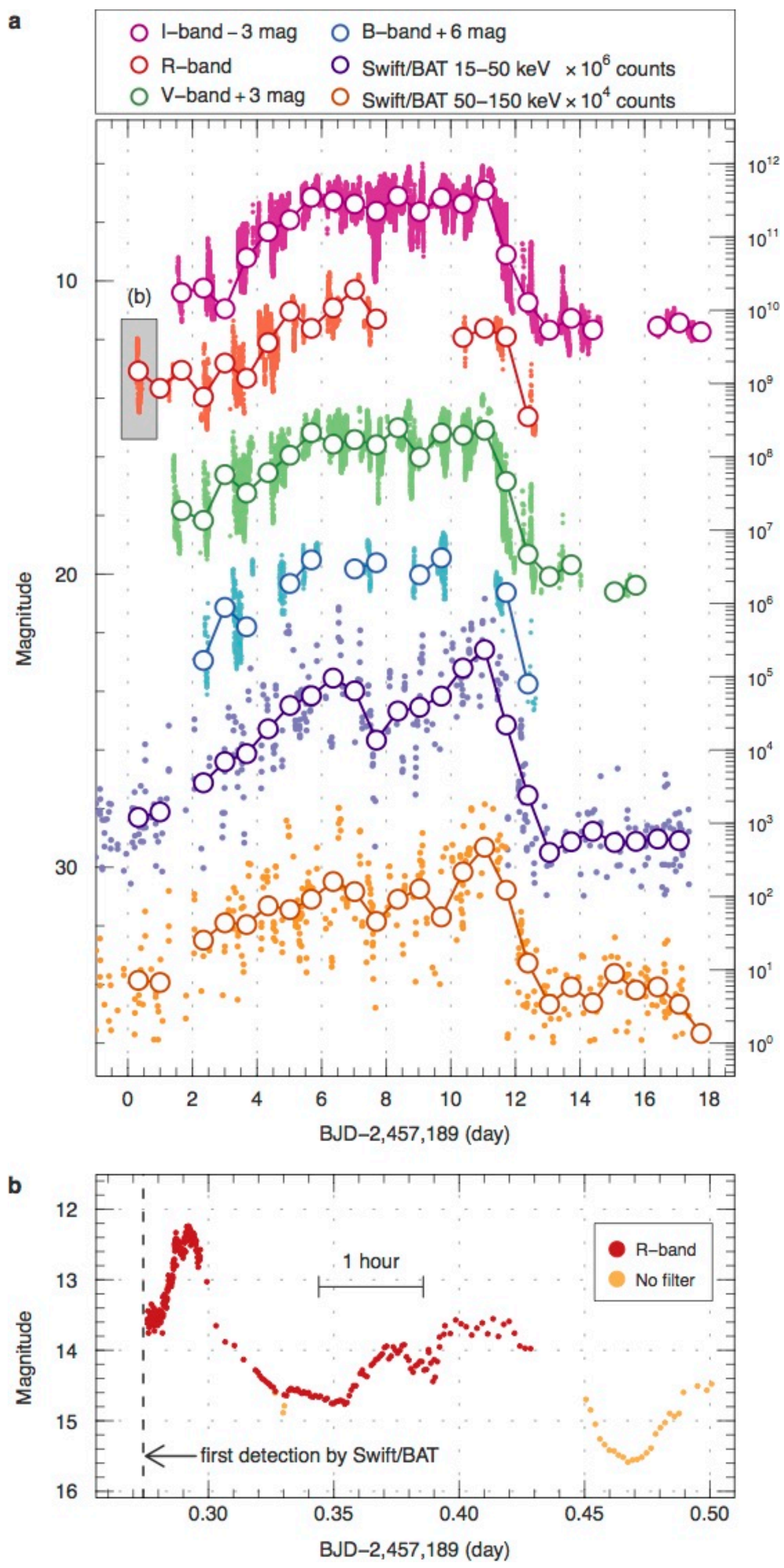

**Extended Data Fig. 1: Optical and X-ray light curves of V404 Cyg during an outburst in 2015 June–July.** Panel a shows overall multi-colour light curves and Swift/BAT light curves. The plotted points are averaged for every 0.67 days. Panel b is an enlarged view of the shaded box in panel a (the first detection of short-term variations). On BJD 2,457,203, the mean magnitude dropped below V =17.0. Superimposed on this rapid fading, the amplitude of variations became progressively smaller and smaller. After BJD 2,457,205, the mean magnitude seemed to be constant, and the outburst virtually ended.

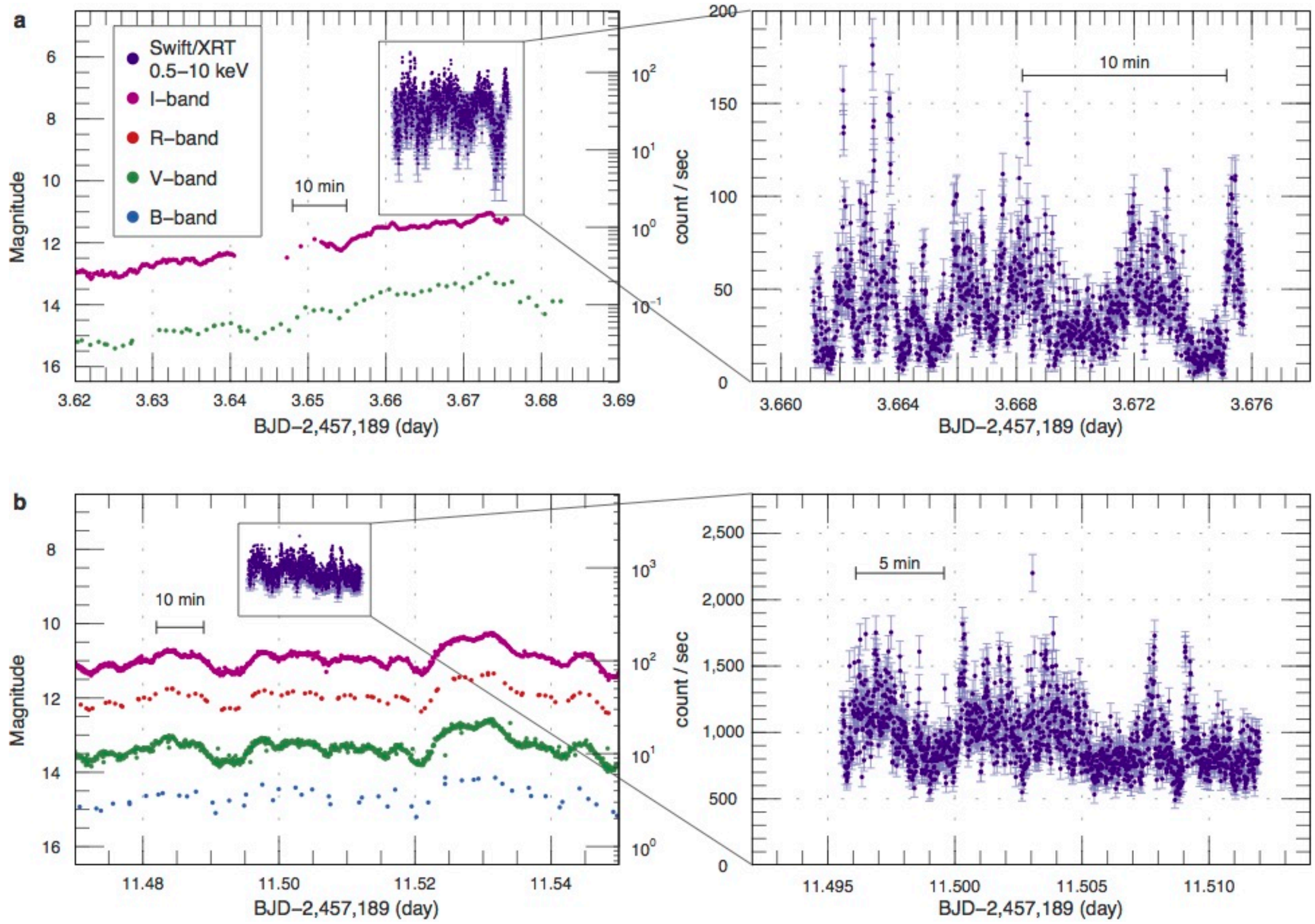

**Extended Data Fig. 2: Additional examples of simultaneous optical and X-ray observations of V404 Cyg in the 2015 outburst except those in Fig. 3.** The left panels of panels a and b represent the correlations on BJD 2,457,192 and BJD 2,457,200, respectively. In the right panels, Swift/XRT light curves in linear scales are shown. The navy blue error bars represent 1σ statistic errors.

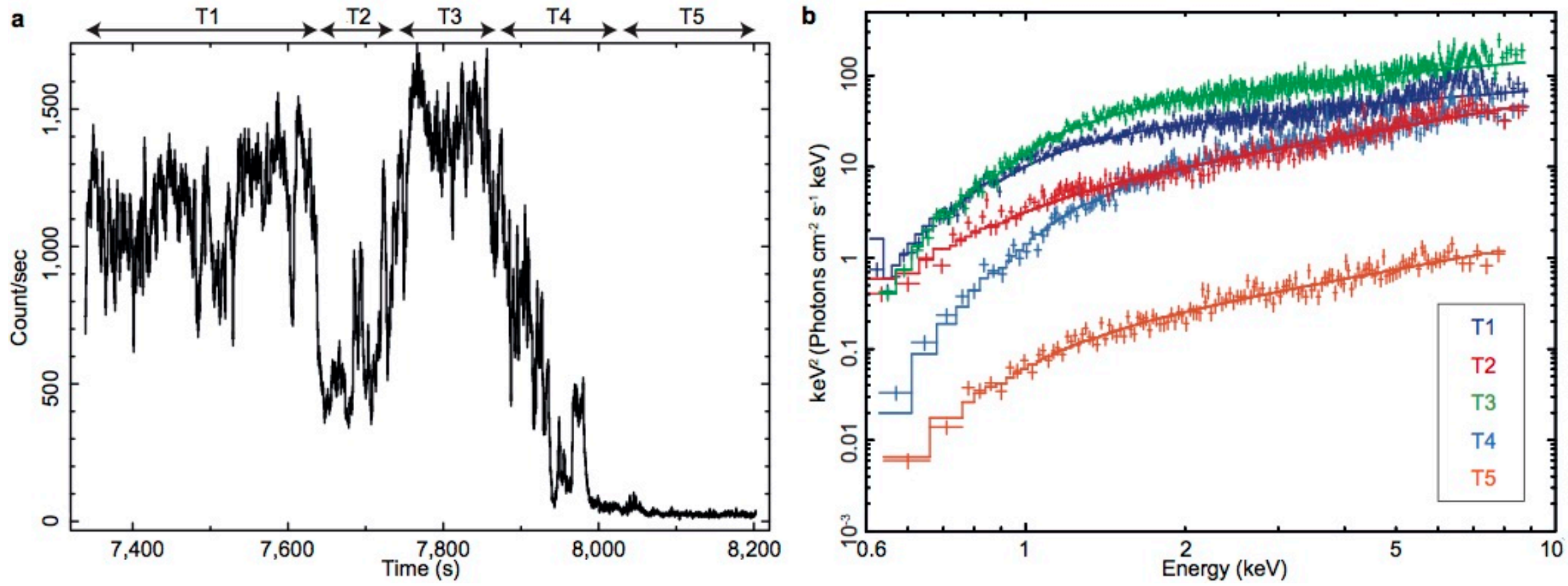

**Extended Data Fig. 3: Example of the soft X-ray light curve and spectra during the dip-type oscillation in the 2015 outburst of V404 Cyg.** (a) The ∼860 s-long Swift/XRT raw light curve (BJD 2,457,194.125–2,457,194.135,ObsID 00031403040) without pile-up correction, same as the X-ray data in Fig. 3a of the main paper. (b) Time-sliced soft X-ray spectra with pile-up correction, in the intervals of T1 to T5 determined in panel a. The exposures of individual spectra are ∼100–300 sec. The error bars represent 1σ statistics errors.

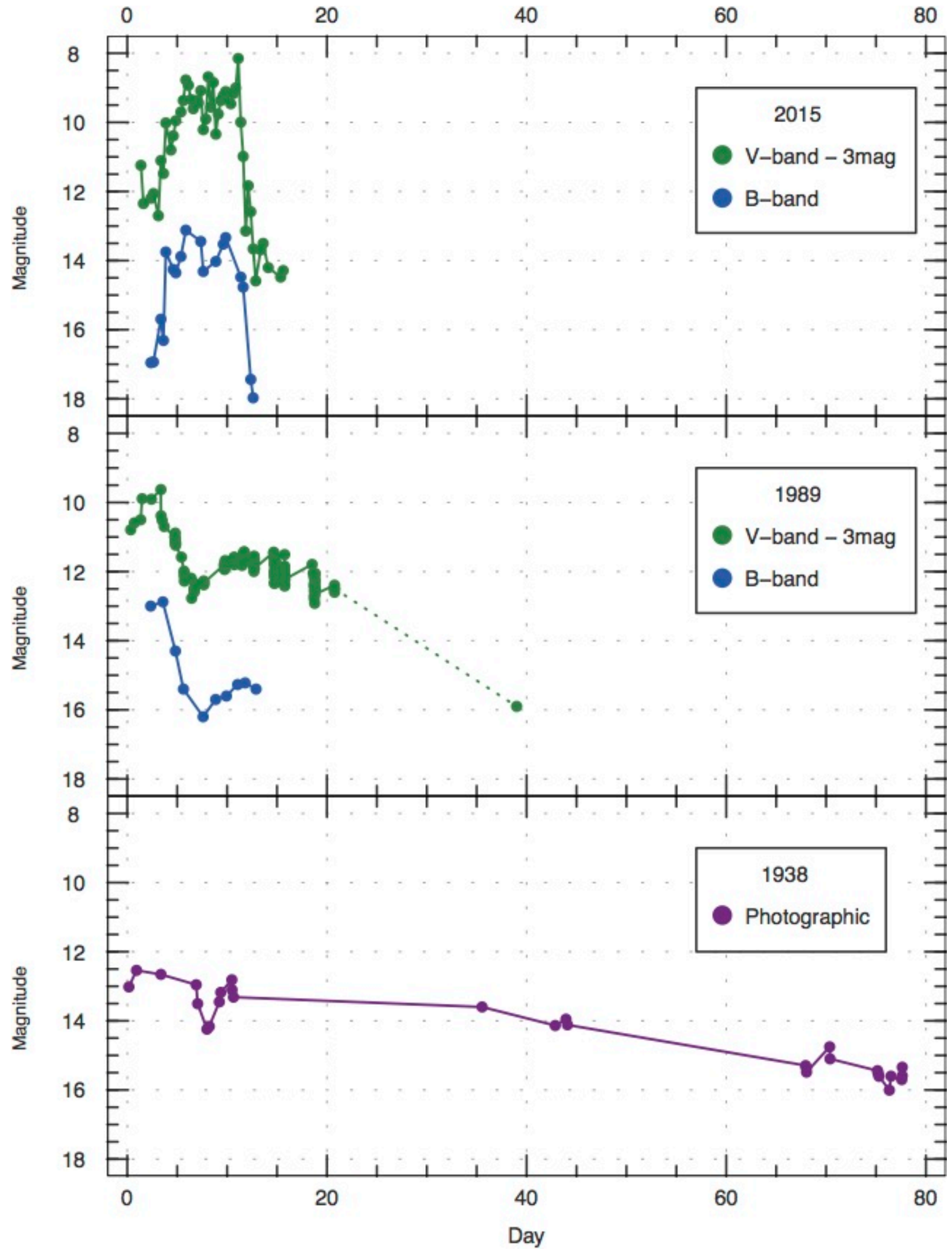

**Extended Data Fig. 4: Comparison of the 1938, 1989 and 2015 outbursts of V404 Cyg.** The horizontal axis represents days BJD−2,429,186, BJD−2,447,673, and BJD−2,457,189, respectively. Photographic magnitudes are approximately the same as B-band.

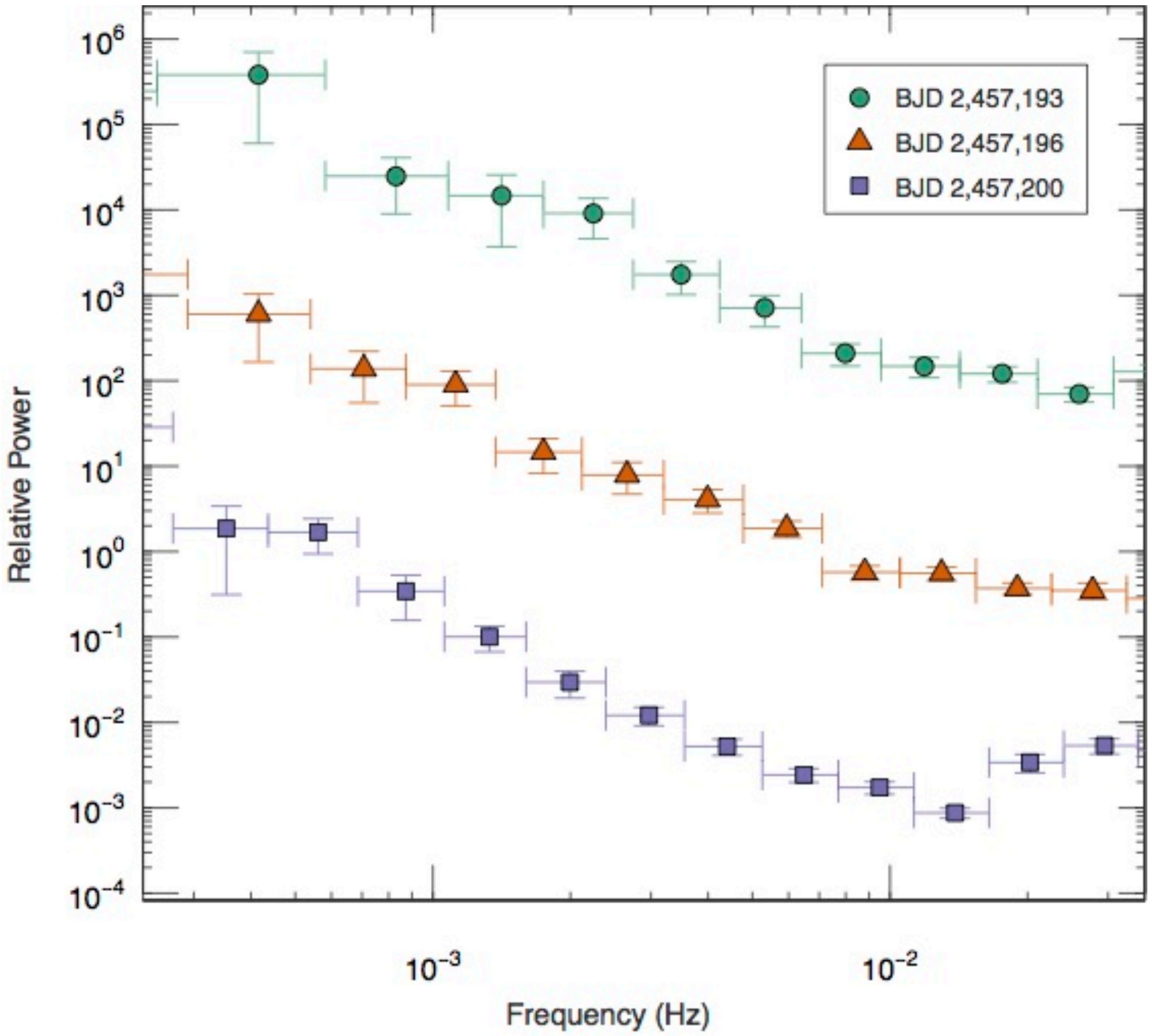

**Extended Data Fig. 5: Power spectral densities of the early stage, the middle stage, and the later stage in the 2015 outburst of V404 Cyg.** Power spectral densities of the fluctuations on BJD 2,457,193 (top, circles), BJD 2,457,196 (middle, triangles), and BJD 2,457,200 (bottom, rectangles). The abscissa and ordinate denote the frequency in Hz units and the power in arbitrary units, respectively. For better visualization, the obtained spectrum is multiplied by $8\times10^{-4}$ on BJD 2,457,196 and by $10^{-4}$ on BJD 2,457,200. The errors are 1σ-corresponding values obtained from relevant chi-square distributions of the power spectra.

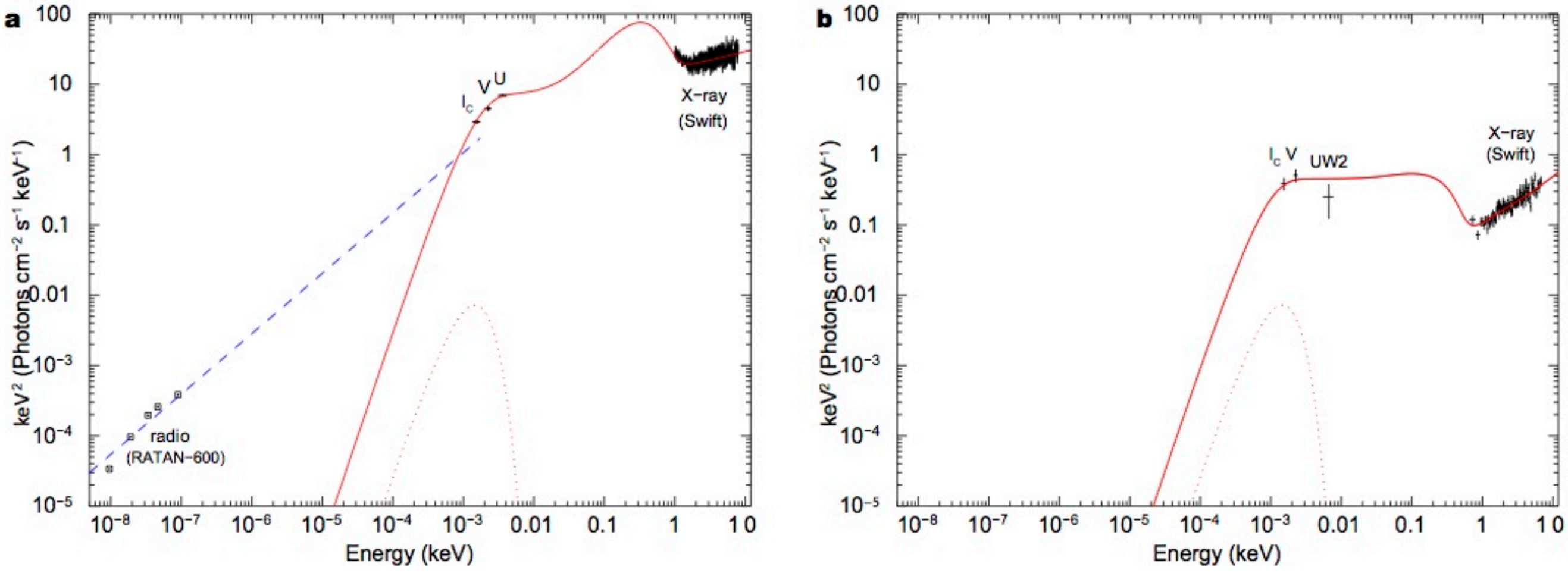

**Extended Data Fig. 6: Simultaneous, extinction-corrected multi-wavelength SEDs of V404Cyg.** The intervals shown are (a) BJD 2,457,199.431–2,457,199.446 and (b) BJD 2,457,191.519–2,457,191.524. The optical (V and $I_C$) fluxes are averaged over the intervals, and the error bars represent their standard errors. The X-ray, U and UW2-band data are obtained with Swift and the errors represent1σ statistic errors. The radio fluxes (open squares) are compiled from the RATAN-600 results atBJD 2,457,199.433[68]. The red solid and dotted lines show the contribution of emissions from the irradiated disc with Comptonisation and from the companion star, respectively. The blue dashed line approximates the radio SED, which is extended to the optical bands for illustrative purposes.